# Review of Energy Transition Policies in Singapore, London, and California

## Online Supplementary Materials for "Data-Driven Prediction and Evaluation on Future Impact of Energy Transition Policies in Smart Regions

Chunmeng Yang[1], Siqi Bu[1], Yi Fan[2*], Wayne Xinwei Wan[3], Ruoheng Wang[1], Aoife Foley[4]

August 2022

## Abstract

The paper reviews the renewable energy development and policies in the three metropolitan cities/regions over recent decades. Depending on the geographic variations in the types and quantities of renewable energy resources and the levels of policymakers' commitment to carbon neutrality, we classify Singapore, London, and California as case studies at the primary, intermediate, and advanced stages of renewable energy transition, respectively.

[1]Department of Electrical Engineering, Hong Kong Polytechnic University. [2]Department of Real Estate, National University of Singapore. [3]Department of Land Economy, University of Cambridge. [4]School of Mechanical and Aerospace Engineering, Queen's University Belfast. [*]indicates corresponding author. E-mails: yi.fan@nus.edu.sg.
Yi Fan acknowledges financial support from MOE AcFR Tier 1 Grant No. R-297-000-145-115.

# Review of Energy Transition Policies in Singapore, London, and California

To promote the transition from fossil energy to renewables while balancing the interests of all stakeholders, various types of policies have been formulated in cities worldwide over the past decade, such as the direct policy, integrating policy, and enabling policy [1]. Direct policies take effects at a strategic level to impact the target renewable energy generation. Particularly, the direct policy can be classified into three main types, namely obligations & certificates, feed-in-tariff, and financial incentives [2]. Obligations refer to the policy instruments that outline the required renewable energy generation, and certificates are provided to acknowledge the compliance with the obligations. The feed-in-tariff policy encourages end-users to generate more renewable energy (e.g., install solar photovoltaic or wind power system at premises), by promising to purchase these energies back at a higher price than the market energy tariff. Financial incentives are normally provided for the large-scale application of new renewable energy technologies. In contrast, integrating policies and enabling policies take effect at the operational level. Specifically, the former provides guidance to the adoption of new renewable energy technologies into the existing energy system and institutional framework, while the latter aims to ensure effective and stable operation of the renewable energy systems. In this paper, we review the renewable energy development and policies in the three metropolitan cities/regions over recent decades. Depending on the geographic variations in the types and quantities of renewable energy resources and the levels of policymakers' commitment to carbon neutrality, we classify Singapore, London, and California as case studies at the primary, intermediate, and advanced stages of renewable energy transition, respectively.

## 1. London

The first city to be reviewed is London, the capital city of the United Kingdom (UK) with a total area of 1,583 square kilometers. According to the UK's Office for National Statistics, the population of London in 2019 is approximately 8.98 million with an annual increasing rate of 0.94%. As one of the world's financial centers, the economy of London is prosperous and robust. The GDP of London in 2018 is GBP 487,145 million, and the nominal and real growth in 2019 is 5.3% and 1.5%. The real economic growth rate is expected to further increase marginally to 1.6% in 2020, before reaching 2.2% in 2021[2]. London is also forecasted to see increases in the number of workforce jobs from 2020 to 2021. Particularly, the GDP of electricity, gas, steam and air conditioning supply sector has been marginally decreasing [4]. The real changes from 2017 to 2019 are -3.6%, -0.1% and -0.51%, respectively.

On the energy supply side, Britain's electricity is currently generated from both the tradition fossil resources like coal and natural gases, and the renewable energy resources like solar, wind and nuclear power. The country's National Grid Electricity System Operator (ESO) collaborates with the other private energy generation companies to balance between energy supply and demand. Electricity from all different generation companies is transmitted solely by the ESO through the Britain's nationwide transmission network. Then the local Distribution Network Operators (DNOs) purchase electricity from the generation companies through transmission network, transmit the electricity to their own local power network, and finally resell the electricity to end-users. In London area, the only local DNO is the UK Power Networks.

[2] This is economic forecast before the COVID-19 pandemic. The forecast of London's real GDP growth is adjusted to -16.8% as in June 2020, but the growth is expected to return to normal figures in 2020 [3].

The UK has a strong green ambition for developing renewable energy, probably due to its affluent natural resources. As the UK is one of the best locations for wind power generation in Europe, the wind farms are developed rapidly and have produced more electricity than coal power plants since 2016. In London, the largest offshore wind farm, the London Array, has 175 turbines in operation currently. Also, the solar energy generation has increased rapidly, due to the government's feed-in tariff schemes and the technology advancements in reducing the cost of photovoltaic panels in recent years. Moreover, technology innovation allows solar energy to be generated at locations other than the traditional ones like building rooftops. In March 2016, more than 23,000 solar panels, with a total capacity of 6.3 MW, are set to float on the surface of Queen Elizabeth II Reservoir.

As a result, over the past two decades, the UK government has implemented a series of policies to encourage the development of renewable resources in different aspects. Table 1 presents the summary of these policies in the UK. The initial strand of policies started in the early 2000s, when the government first invested in the R&D of the renewable energy related technologies. Grants and subsidies were provided, which a focus on the wind and marine energy research. Financial incentives were also provided for large-scale demonstration of renewable energy production. In 2002, the Renewable Obligation (RO) was introduced, which was among the UK's first comprehensive renewable energy development scheme with various policy instruments. It requires the UK electricity suppliers to source an increasing proportion of electricity from renewable sources, supported through the Feed-in-Tariffs, and it also establishes the various certifications to acknowledge the providers' achievements. More action plans are progressively introduced in the following years, and the scope is also expanded to other resources such as solar and nuclear powers. As a result, the production of renewable energies in London as significantly increased from around 400 GWh in 2004 to 1,100 GWh in 2018[5].

More ambitious targets are also set by London's government to tackle the future climate change. The objective is to supply 15% of London's energy by renewable energy by 2030. On 27th Jun 2019, UK became the first major economy to commit a net-zero emission goal by the year 2050, which means that the country will reduce at least 80% of the carbon emission compared with that in 1990. To achieve the zero-carbon city target for London, on 13th Jan 2020, a public-owned company called London Power is established by the government, which provides electricity solely generated from clean energy resources like solar, wind and hydro power. Some novel market programs are also introduced, such as "Solar Together London", which encourages the residents to install solar panels on their homes by arranging wholesale purchases at an affordable price.

**Table 31. Summary of Renewable Energy Policies in London**

| Renewable Energy Policy | Start Year | End Year | Policy Types |
|---|---|---|---|
| Research & Development and Demonstration of Wave and Tidal-stream Technologies | 2000 | / | Fiscal/financial incentives |
| Energy Technology Programme | 2002 | 2013 | Research; Development and deployment |
| Large-scale PV Demonstration Project | 2002 | 2007 | Regulatory instruments; Auditing; Fiscal/financial incentives; Grants/subsidy; Research, development and deployment; Experimental development; Demonstration; Enabling Legislation; Monitoring |
| Offshore Wind Capital Grants Scheme | 2002 | 2004 | Fiscal/financial incentives; Grants/subsidy |
| Renewables Obligation (RO) | 2002 | / | Regulatory instruments; Codes and standards; Utility obligations; Performance labels; Comparison label; Market-based instruments; Green certificates; Obligation schemes; Fiscal/financial incentives; Enabling Legislation; Monitoring |
| Renewable Energy Guarantees of Origin (REGOs) | 2003 | / | Regulatory instruments; Codes and standards; Performance labels; Comparison label; GHG emissions trading; Enabling Legislation; Monitoring |
| Marine Research Development Fund (MRDF) | 2005 | 2011 | Research, development and deployment; Demonstration; Deployment |
| Energy Technologies Institute | 2007 | / | Strategic planning; Voluntary approaches; Negotiated agreements (public-private sector); Research, development and deployment; Framework Policy |
| Environmental Transformation Fund (ETF) | 2007 | 2011 | Economic instruments; Fiscal/financial incentives; Grants/subsidy |
| Climate Change Act 2008 | 2008 | / | Strategic planning; Regulatory instruments; Utility obligations; Economic instruments; Market-based instruments; GHG emissions trading; Obligation schemes |
| Energy Act 2008 | 2008 | / | Strategic planning |
| Planning and Energy Act 2008 (England and Wales) | 2008 | / | Regulatory instruments |
| Renewable Energy Strategy 2009 | 2009 | / | Strategic planning; Regulatory instruments; Fiscal/financial incentives; Feed-in tariffs/premiums; Research, development and deployment; Experimental development; Deployment |
| Feed-in Tariffs for renewable electricity for PV and non-PV technologies | 2010 | 2019 | Fiscal/financial incentives; Feed-in tariffs/premiums |
| National Renewable Energy Action Plan (NREAP) | 2010 | / | Strategic planning |
| Energy White Paper 2011 | 2011 | 2015 | Strategic planning; Regulatory instruments; Codes and standards; Fiscal/financial incentives; Feed-in tariffs/premiums |
| National Energy Efficiency Database | 2011 | / | Regulatory instruments; Information and education |
| Decentralised Energy Capacity Study | 2011 | 2011 | Strategic planning; Research, development and deployment; Framework Policy |
| EU Directive (2012/27/EU) | 2013 | / | Strategic planning; Target |
| Electricity Market Reform (EMR) | 2013 | / | Regulatory instruments |
| Green Deal | 2013 | 2019 | Economic instruments; Direct investment; Fiscal/financial incentives; Grants/subsidy; Loan / debt finance |
| Contract for Difference (CfD) | 2014 | / | Fiscal/financial incentives; Feed-in tariffs/premiums |
| UK Clean Growth | 2017 | / | Strategic planning |
| Strategy | | | |
| Climate Change Agreements | 2017 | / | Economic instruments; Fiscal/financial incentives; Tax relief |
| London Community Energy Fund | 2017 | / | Fiscal/financial incentives; Development and deployment |
| Mayor of London Energy Efficiency Fund (MEEF) | 2018 | / | Market-based instruments; Fiscal/financial incentives |
| Solar Action Plan | 2018 | / | Demonstration; Research, development and deployment; Framework Policy; Voluntary approaches |
| Solar Together London | 2018 | / | Fiscal/financial incentives; Feed-in tariffs/premiums; Market-based instruments |
| London Power | 2020 | / | Experimental development; Demonstration |

## 2. California

Owing to data limitation in obtaining accurate city-level information in the United States (US), in the second part of the review we focus on the development of renewable energy in the entire state of California. California locates on the west coast of the US with a total area of 423,970 square kilometers, which is the third-largest state by area in the country. According to California's Department of Finance, the latest population in California is approximately 40 million, with an annual growth rate of 0.35% estimated in 2019. As one of the most developed states in the US and the largest sub-national economy of the world, California has a total GDP of USD 3.2 trillion in 2019.

Since the 1990s, the government of California has started to reorganise its electricity market to introduce more competition and to reduce the overall cost for end-users. In the current market structure, electricity is generated by private power generation companies with their self-owned generators and is transmitted by three major transmission companies. At the regional level, different utility distribution companies purchase electricity in a wholesale market and a company named California Independent System Operator (CAISO) is in charge of dispatching. End-use consumers have options to choose the utility distribution companies, generating competition over retail electricity price. The Federal Energy Regulatory Commission (FERC) oversees the market; however, California utilises the market to determine the supply and source of energy. As the public electricity retailers do not own any power plants, in some cases they need to purchase electricity from other states to meet the demand. As in 2018, almost one-third of electricity supply to California is provided from outside of the state.

California is in a leading position of renewable energy development throughout the country. For the electricity generated within California, renewable energy constitutes around a half of the net electricity generation, in which the nuclear and hydroelectric power are two major sources. According to the Energy Information Administration (EIA), in February 2020, the natural gas-fired plants generate around 6,000 GWh, while the generations for nuclear power, hydroelectric power and other renewable resources are 1,500 GWh, 1,100 GWh, and 4,500 GWh, respectively. However, the production of renewable energy from these two sources in future are in great uncertainty. First, after the permanent shutdown of the San Onofre Nuclear Power Station in year 2013, only two nuclear reactors are still under operation in California, and these two nuclear reactors are also planned to be decommissioned between 2024 and 2025. No more nuclear power stations are planned to operate in the coming few years. Second, although California has the second-largest conventional hydroelectric generating capacity in USA, the hydropower production is not stable and strongly depends on the amount of rainfall.

To tackle these challenges in future development of renewable energy, the government maintains its advancement in green ambition and has progressively shifted its focus of renewable energy policies in the past decades from expanding the overall capacity to developing new renewable energy resources. Table 2 presents the major renewable energy policies in California. The state's first large-scale investment in renewable energy production was first introduced in 1980s, when the Federal Public Utility Regulatory Policies Act (PURPA) required all power companies to purchase power generated by qualified renewable resources. One of most effective policies in expanding the renewable energy capacity is the introduction of Renewable Portfolio Standards (PRS), which requires the electricity retailers to have a fixed percentage of power generated from renewable resources. The percentage was initially set to be 20% in 2002, and it was increased to 33% in 2020. Started from 2002, several new incentives are introduced in the state to further diversify the renewable energy resource, aiming to create a balanced and reliable mixture of renewable energy production in California. Utilising the natural resources and human capital in the Silicon Valley [6–8], new renewable energy technologies that are adopted in California by 2018 include geothermal, solar, wind, landfill gas and biomass.

In summary, California has expressed a strong green ambition to further boost the development of renewable energy generation while maintaining the prosperity of its economy. One of its ambitious targets is to significantly reduce the emission of greenhouse gases (GHG). The California Senate Bill 350, which was signed in 2017, sets the goals of reducing petroleum emission by 50% and increasing the percentage of renewable energy in load-serving electricity to at least 50% by 2030. In the same legislative session, the Senate Bill 32 sets the GHG emission goal in 2030 to be 40% lower than the emission level in 1990.

**Table 2. Summary of Renewable Energy Policies in California**

| Renewable Energy Related Policy | Start Year | End Year | Policy Types |
|---|---|---|---|
| Solar Photovoltaic Energy Research, Development and Demonstration Act | 1978 | 1978 | Research; Development and deployment |
| Public Utility Regulatory Policies Act (PURPA) | 1978 | / | Economic instruments; Market-based instruments |
| Wind Energy Systems Act | 1980 | 1980 | Research; Development and deployment |
| State-level Renewable Portfolio Standards (RPS) | 1983 | / | Regulatory instruments; Codes and standards; Obligation schemes |
| Modified Accelerated Cost Recovery System (MACRS) | 1986 | / | Tax relief |
| Renewable Energy Production Incentive (REPI) | 1992 | 2005 | Regulatory instruments; Tax relief |
| Federal Business Investment Tax Credit (ITC) | 1992 | / | Tax relief |
| Renewable Electricity Production Tax Credit (PTC) | 1993 | / | Strategic planning; Tax relief |
| Tribal Energy Program | 1994 | / | Information and education; Advice/aid in implementation; Fiscal/financial incentives |
| Federal Utility Partnership Working Group (FUPWG) | 1994 | / | Regulatory instruments; Information and education |
| State Energy Program | 1996 | / | Regulatory instruments; Information and education; Advice/aid in implementation; Fiscal/financial incentives |
| WIND Exchange | 1999 | / | Information and education; Advice/aid in implementation; Fiscal/financial incentives; Research; Development and deployment |
| San Francisco Solar Energy Incentive Program | 2001 | / | Fiscal/financial incentives |
| Green Power Partnership | 2001 | / | Strategic Planning; Information and education; Advice/aid in implementation |
| Renewable Portfolio Standard -California | 2002 | / | Regulatory instruments; Codes and standards; Obligation schemes |
| Woody Biomass Utilisation Initiative | 2003 | / | Information and education; Fiscal/financial incentives |
| State Utility Commission Assistance | 2005 | / | Regulatory instruments; Information and education; Advice/aid in implementation |
| Interconnection Standards for Small Generators | 2005 | / | Regulatory instruments; Codes and standards |
| Clean Energy-Environment State Partnership Program | 2005 | / | Strategic Planning; Regulatory instruments; Information and education; Advice/aid in implementation |
| Solar America | 2006 | 2015 | Strategic Planning; Information and education; Advice/aid in implementation; Research; Development and deployment |
| Section 1703/1705 Loan Guarantee Program | 2006 | / | Fiscal/financial incentives; Loan / debt finance |
| Community Renewable Energy Deployment Grants | 2007 | / | Direct investment; Funds to sub-national governments; Fiscal/financial incentives |
| California Solar Initiative | 2007 | / | Fiscal/financial incentives |
| Wind power development: MOU between DOE and wind companies | 2008 | 2010 | Research; Development and deployment |
| Wind Energy Technologies Office (WETO) | 2008 | / | Regulatory instruments; Information and education; Advice/aid in implementation; Research; Development and deployment |
| Advanced Solar PV development: Solar America Initiative | 2008 | / | Research; Development and deployment |
| Rapid Deployment of Renewable Energy and Electric Power Transmission Projects | 2009 | 2011 | Fiscal/financial incentives |
| Final Rule on Renewable Energy and Alternate Uses of Existing Facilities on the Outer Continental Shelf | 2009 | 2012 | Regulatory instruments; Research; Development and deployment |
| Bureau of Land Management Renewable Energy Resources | 2009 | / | Information and education; Fiscal/financial incentives |
| American Recovery and Reinvestment Act: | 2009 | / | Direct investment; |
| Appropriations for Clean Energy | | | Fiscal/financial incentives;<br>Research;<br>Development and deployment |
| Smart from the Start Initiative | 2010 | / | Regulatory instruments;<br>Codes and standards |
| Renewable Feed-In Tariff (FIT) Program (CPUC) | 2013 | / | Feed-in Tariff |
| Renew300 Federal Renewable Energy Target | 2015 | / | Strategic planning |
| Water Power Technologies Office (WPTO) | 2016 | / | Information and education;<br>Research;<br>Development and deployment |
| Solar Energy Technologies Office/SunShot (SETO) | 2016 | / | Information and education;<br>Research;<br>Development and deployment |
| Hydroelectric Production Incentive Program | 2016 | / | Economic instruments |

## 3. Singapore

Singapore locates at the southernmost point of the Malay Peninsula, and it is one of the world's leading international financial centers. As a city-state, the country has a total area of 725.1 square kilometers. The total resident population is 5.7 million and the annual growth rate is 1.2% in 2019. The total GDP of Singapore in 2019 is SGD 507.6 billion (equivalent to approximately USD 371 billion), with is a real growth of 0.7% from 2018. According to the Word Bank, the GDP per capital of Singapore ranks the $9^{th}$ in the world, which is slightly lower than the US but higher than the UK.

Like California, the Singaporean government in the recent years has introduced more competition in the retail market to lower the price of electricity. Before the 1990s, Singapore's electricity power generation, transmission, and distribution processes are owned and managed by the state-owned enterprises. In 1998, the country started the reform of electricity market, as it was the first country that adopted wholesale electricity market in Southeast Asia. In order to promote effective competition, the Energy Market Authority (EMA) was established in April 2001 to oversee the market. In 2018, the Open Electric Market policy is introduced, aiming to further increase the flexibility for end-users. Specifically, customers can choose the price plan from various electricity retailors to meet their needs, while the electricity supply is still centrally transmitted under the management of the grid company. As in 2019, there are over 21 licensed electricity retailers that are allowed to sell electricity to consumers.

Solar power is the major source of renewable energy in Singapore, but in general, renewable energy only constitutes a small proportion of the country's energy consumption. As an island city with limited energy resources, more than 90% of electricity in Singapore is generated by natural gas, and the natural gas is all imported from other countries (72% from Indonesia and Malaysia through pipelines, and 5% is received and stored in the country's own liquefied natural gas terminal). The low production of renewable energy is mainly due to the restriction from natural resources. The average wind speed in Singapore is only around 2 meter per second (m/s), which is smaller than the minimum required wind speed (4.5m/s) for the standard commercial wind turbines. The hydro resources and geothermal energy are also unavailable in Singapore. As for the solar power, Singapore locates near the equator and the average annual solar irradiance in the country is around 1580 kWh per square meters per year, which is around 50% stronger than other countries in the temperate zone. Thus, solar energy is the most potential renewable energy to meet the growing energy demand in Singapore. However, Singapore is also a densely populated city-state, which lacks sufficient areas to install photovoltaic panels.

As a result, the majority of renewable energy related policies in Singapore focus on the development of solar power, especially for funding the research in improving the power generation efficiency within limited areas. Table 3 summarises these polices in Singapore. For instance, the Clean Energy Programme Office (CEPO) was established in 2007 to provide the SGD 50 million funding to support R&Ds in solar photovoltaics. Also, the country's housing development board is given a total funding of S$31 million to install solar panels on the rooftops of residential blocks in 30 public housing precincts by 2015. More recently, Singapore successfully developed and installed 10 different floating support structures for PV systems that were constructed by both local and overseas companies on the Tengeh Reservoir in 2016

to determine the most suitable system for Singapore [9]. Building on the results of the test bed, the Public Utilities Board (the nation’s water agency) is now exploring the feasibility of deploying a 50 MWh floating solar PV system at the Tengeh Reservoir. The amount of energy generated from such a system could potentially power about 12,500 average households in Singapore. In total, with the effect of these policies, the annual production of renewable energy in Singapore has increased from 479 GWh in 2004 to 918 GWh in 2015 [10].

In comparison with London and California, Singapore is not expected to aggressively increase the percentage of renewable energy in its total energy consumption due to the limit of resources, but it still holds a strong ambition to develop clean energy in the county. One major reason is, as a tropic island, Singapore concerns about the effect of greenhouse emission and global warming. The most remarkable greenhouse gas in Singapore is carbon dioxide, majorly created by the combusting of fuel resources to satisfy the needs of power used every day. Singapore government aims to reduce the intensity of carbon emissions in 2030 by 36% in comparison to the level of 2005. Also, as part of the country’s strategic plan, it aims to rely less on the import of fossil energy in the future [11], so more investment in developing cleaning energies like solar power is expected.

**Table 3. Summary of Renewable Energy Policies in Singapore**

| Renewable Energy Policy | Start Year | End Year | Policy Types |
|---|---|---|---|
| Joint Research with Tertiary Institutions | 1991 | / | Research; Development and deployment |
| Tax Incentive for Energy-saving Equipment | 1996 | / | Strategic planning; Tax relief |
| Innovation for Environmental Sustainability (IES) Fund | 2001 | / | Research; Development and deployment |
| Energy Innovation programme office (EIPO) Singapore | 2007 | / | Strategic planning |
| Clean Energy Research Programme (CERP) | 2007 | / | Research; Development and deployment |
| Solar Pilot/Test-bedding programmes | 2007 | 2014 | Strategic planning |
| Solar Energy Research Institute of Singapore (SERIS) | 2008 | / | Research; Development and deployment |
| Energy National Innovation Challenge (NIC) Singapore | 2011 | / | Strategic planning; Economic instruments; Research, development and deployment; Enabling Legislation |
| Floating PV Pilot | 2011 | 2017 | Research, development and deployment; Demonstration |
| Energy Innovation Research Programme (EIRP) | 2012 | / | Fiscal/financial incentives; Grants/subsidy; Research, development and deployment |
| Renewable Energy Certificate Marketplace (SG Group) | 2018 | / | Obligations and Certificates |